\documentclass[amssymb,amsmath,preprint]{revtex4-1}
\usepackage[colorlinks]{hyperref}
\usepackage{graphicx}

\usepackage{gensymb}

\begin{document}
\title{Thermal leptogenesis in anisotropic cosmology}

\author{Mehran Dehpour}
\email{m.dehpour@mail.sbu.ac.ir}
\affiliation{Department of Physics, Shahid Beheshti University,
PO Box 19839-63113, Tehran, Iran}

\begin{abstract}
    There is no evidence that the universe must have been homogeneous and isotropic before the big bang nucleosynthesis. The Bianchi type-I cosmology is the simplest homogeneous but anisotropic cosmology. In this work, we investigate thermal leptogenesis, as a baryogenesis scenario, in the Bianchi type-I cosmology. Our results show that for specific values of the anisotropy, the modified thermal leptogenesis generated more baryon asymmetry than the standard one. In this way, anisotropy can help to achieve low-scale leptogenesis.
\end{abstract}

\maketitle

\section{Introduction}
\label{sec:intro}
Observations confirm that all we know in the universe is made of baryons, and there is almost no antibaryon in structures. The baryon asymmetry of universe is defined as
\begin{align}
    Y^{\rm obs}_{B} \equiv \left. \frac{n_B - \overline{n}_{B}}{s} \right|_0 = (8.73 \pm 0.35) \times 10^{-11},
\end{align}
where $n_B$, $\overline{n}_{B}$, and $s$ are the number densities of baryons, antibaryons, and entropy respectively. Also, the subscript $0$ denotes the present time. The value of the baryon asymmetry is determined through Big Bang Nucleosynthesis (BBN), Cosmic Microwave Background (CMB), and Large-Scale Structure (LSS) observations at a confidence level $95\%$ \cite{Simha:2008zj}.

If we consider that the universe was produced initially without baryon asymmetry or that the initial baryon asymmetry was washed by inflation, we expect that for every baryon there is an antibaryon, which we cannot find in the universe. Scenarios that attempt to answer the question of where the antiparticles are going are called baryogenesis. According to Sakharov, all of these scenarios depend on three essential conditions \cite{Sakharov:1967dj}: violation of baryon number conservation, C and CP violation, and the presence of out-of-equilibrium dynamics. Although these conditions exist in the standard model, no identified framework can produce the desired baryon asymmetry \cite{Gavela:1994ds,Gavela:1994dt}. Therefore, it is necessary to search beyond the standard model physics (BSM) for suitable baryogenesis scenarios.

Thermal leptogenesis is a successful baryogenesis BSM scenario based on the extension of the standard model by adding at least two right-handed neutrinos (RHNs) \cite{Fukugita:1986hr}. These new particles were introduced using the seesaw mechanism \cite{Mohapatra:1980yp,Yanagida:1979as,Glashow:1979nm,Gell-Mann:1979vob,Minkowski:1977sc}. Decaying RHNs through the Yukawa channel leads to a new source of out-of-equilibrium CP violation that provides a new framework to satisfy three Sakharov conditions and generate the observed baryon asymmetry of the universe.
The problem of thermal leptogenesis is the huge lower bound on the RHN masses \cite{Davidson:2002qv}. This bound can lead to gravitino overproduction in conflict within supersymmetric models \cite{Kawasaki:2008qe,Rychkov:2007uq,Kawasaki:1994af,Khlopov:1984pf,Weinberg:1982zq}. Furthermore, the huge mass required for thermal leptogenesis leads to its phenomenological untestability because of the presence of an inaccessible energy part in the present experiments. Modern developments in thermal leptogenesis have attempted to solve these problems. One branch of these studies uses nonstandard cosmologies such as Refs. \cite{Chen:2019etb,Dutta:2018zkg,PhysRevD.90.064050}.
Similarly, nonstandard cosmologies have also been regarded in other baryogenesis scenarios, such as gravitational baryogenesis \cite{Bhattacharjee:2021jwm,Bhattacharjee_2020,Bhattacharjee:2020jfk,Baffou:2018hpe,Atazadeh:2018xjo,Fukushima:2016wyz,Oikonomou:2016jjh,Odintsov:2016hgc,Saaidi:2010ey} and electroweak baryogenesis \cite{Aliferis:2020dxr,Barenboim:2012nh}.

In this study, we focus on using a nonstandard cosmology type of development by neglecting the isotropic cosmological principle in the early universe. As we do not have signatures of isotropics before the BBN, this assumption is rational. Moreover, the increasing accuracy of recent observations has created tensions in standard cosmology \cite{Verde:2019ivm,DiValentino:2020zio,DiValentino:2021izs}. Several attempts have been made to reduce these tensions \cite{Abdalla:2022yfr,Perivolaropoulos:2021jda}. This problem makes the cosmological principles of the isotropic and homogeneous universe suspicious \cite{Colgain:2022tql,Colgain:2022rxy,Krishnan:2021dyb,Krishnan:2021jmh}. To review this topic, refer to Ref. \cite{Aluri:2022hzs}. Thus, we have recently witnessed an improvement in anisotropic cosmology models. The most famous class is the Bianchi cosmology \cite{Ellis:1968vb}. Among these, Bianchi type I (BI) is the simplest, which we focus on in this study. More information about BI can be found in Refs. \cite{delliou_anisotropic_2020,Russell:2013oda,jacobs1969bianchi}. Here, we found the effect of the anisotropy of universe on thermal leptogenesis with three RHNs. In this way, we showed that the Hubble expansion rate was modified, which caused the decay parameter to increase with a moving forward, reducing the washout strength, and decreasing the produced RHN abundance. As mentioned later, the effect of changing the decay parameter rising time on the resulting baryon asymmetry is negligible. In addition, it is clear that a decrease in washout results in a greater baryon asymmetry. In contrast, the production of RHN with anisotropic cosmology is reduced, resulting in a reduction in the amount of CP violation through the decay of RHN. As a result of this competition, anisotropic universe generates more baryon asymmetry for specific amounts of anisotropy.

This paper is organized as follows. In Sect. \ref{sec:an-isotropic-cosmology}, we briefly introduce anisotropic BI cosmology. In Sect. \ref{sec:modified-leptogenesis}, we provide a detailed review of thermal leptogenesis and impose the anisotropy effect on it. In Sect. \ref{sec:results}, we introduce the parameters of the model and extract our results by numerically solving the Boltzmann equations. Finally, in Sect. \ref{sec:conclu}, we discuss about the results of this work and provide some proposals.

\section{Anisotropic Bianchi type-I cosmology}
\label{sec:an-isotropic-cosmology}
Standard cosmology was formulated using the Friedmann-Lemaitre-Robertson-Walker (FLRW) metric. In the FLRW metric, we assume the space part to be flat and consider cosmological principles that state that the universe should be homogeneous and isotropic on large scales.  However, Bianchi cosmology is an alternative that neglects the isotropy. The simplest example of Bianchi universes is expressed by BI metric \cite{Ellis:1968vb,delliou_anisotropic_2020,Russell:2013oda,jacobs1969bianchi}
\begin{align}
    ds^2 = -dt^2 + a_1^2(t) dx^2 + a_2^2(t) dy^2 + a_3^2(t) dz^2,
    \label{eq:BI-metric}
\end{align}
where $a_i$ are the directional scale factors and the directional Hubble rates are given by $H_i=\dot{a_i}/a_i$.

As we are interested in the early universe, the generalized Friedmann equation in the presence of radiation energy density $\epsilon_r \propto a^{-4}$ is given by
\begin{align}
    H^2 = \frac{8 \pi G}{3} \epsilon_r + \frac{1}{3} \sigma^2,
    \label{eq:Friedmann-Eq.}
\end{align}
where the effective scale factor and Hubble rate are
\begin{align}
    a \equiv \left(a_1 a_2 a_3\right)^{1/3}, \quad H \equiv\dot{a}/a = \frac{1}{3} \left(H_1+H_2+H_3\right).
    \label{eq:Hubble-definition}
\end{align}
In the Friedmann equation, the anisotropy of universe is encoded in the square of the shear scalar which is defined as
\begin{align}
    \sigma^2 \equiv \frac{1}{6} \left[\left(H_1-H_2\right)^2+\left(H_2-H_3\right)^2+\left(H_3-H_1\right)^2\right].
\end{align}
Using the useful relation $\dot{H}_i - \dot{H}_j = -3 H \left(H_i - H_j\right)$ which is equivalent to $H_i - H_j \propto a^{-3}$, one can obtain the square of the shear scalar dependence on the effective scale factor $\sigma^2 \propto a^{-6}$. As a result, the square of the shear scalar decreases more rapidly than the energy density of the radiation.

The temperature at which $8 \pi G \epsilon_r=\sigma^2$ is defined as $T_e$. For $T\gg T_e$ the universe is shear-dominated: $H \propto a^{-3}$ and $a \propto t^{1/3}$ then $H = 1/3t$; for $T \ll T_e$ the universe is radiation-dominated: $H \propto a^{-2}$ and $a \propto t^{1/2}$ then $H = 1/2t$. 
Therefore, we can interpret the temperature $T_e$ as the size of the anisotropy; lower values of $T_e$ correspond to larger anisotropy at a fixed temperature, and vice versa.
The square of the shear scalar can be written in terms of the radiation energy density; then the modified Hubble expansion rate can be written as \cite{PhysRevD.42.3310}
\begin{align}
    H=\frac{1.66}{M_{Pl}} (g_{\star})^{1/2} T^2 \sqrt{1+\frac{g_{\star}}{g^e_{\star}}\frac{T^2}{T^2_e}},
    \label{eq:Hubble}
\end{align}
where $M_{\rm Pl} = 1.22 \times 10^{19}$ is the Planck mass and $g_{\star}$ and $g_{\star}^e$ are the effective degrees of freedom for the energy density at $T$ and $T_e$, respectively. Note that, in the limit of $T_e \to \infty$ Eq.\ (\ref{eq:Hubble}) restore the default Hubble expansion rate. Because we did not observe any signature of anisotropy in the BBN, we believe that anisotropy did not affect it. So, we have the constraint $T_e\gg 2.5\ \rm MeV$ \cite{PhysRevD.42.3310}.

\section{Modified thermal leptogenesis}
\label{sec:modified-leptogenesis}
Thermal leptogenesis is based on the concept of introducing RHNs that interact with standard model particles via Yukawa interactions and gravity. These heavy sterile particles can be created through thermal mechanisms in the early universe, similar to other standard model particles. They can violate CP in their out-of-equilibrium decay through the Yukawa channel if the Yukawa coupling constants are complex, as shown in Eq.\ (\ref{eq:CP-parameter}). This new CP violation source leads to the production of the asymmetry. This asymmetry is communicated from singlet neutrinos to ordinary leptons through their Yukawa couplings. The lepton asymmetry is then reprocessed into baryon asymmetry by the electroweak sphalerons. 

Here, we consider a simple model in which three RHNs exist, but only the lightest of them, $N_1$, can participate in Yukawa interactions. The reactions involving $N_1$ can be described as
\begin{align}
    N_1 \rightleftarrows \bar{\phi} l,
    \label{eq:decay}\\
    N_1 \rightleftarrows \phi \bar{l}.
    \label{eq:decay-anti}
\end{align}
To quantify this scenario, one can first calculate the tree-level decay rates by definition of $\Gamma_1 \equiv \Gamma(N_1 \to \bar{\phi} l)$, and $\overline{\Gamma}_1 \equiv \Gamma(N_1 \to \phi \bar{l})$ as \cite{Davidson:2008bu}
\begin{align}
    \Gamma_1= \overline{\Gamma}_1 = \frac{M_1}{16 \pi} (yy^{\dagger})_{11},
    \label{eq:decay-rate}
\end{align}
where $M_1$ is the mass of $N_1$ and $y$ is the Yukawa coupling matrix. It should be noted that the rates of these decays are lower than the Hubble rate at high temperatures. We write the thermal average of the decay rates which are obtained in Eq.\ (\ref{eq:decay-rate}) as
\begin{align}
    \langle \Gamma_{1} \rangle = \langle \overline{\Gamma}_1 \rangle =  \frac{K_1(z)}{K_2(z)} \frac{M_1}{16 \pi} (yy^{\dagger})_{11},
    \label{eq:averaged-decay-rate}
\end{align}
where $z\equiv M_1/T$ is a dimensionless parameter and $K_n(z)$ is the $n$th order of the second kind of modified Bessel function.

When the temperature is lower than the mass of the lightest RHN, $M_1$, the reactions described in Eqs.\ (\ref{eq:decay}) and (\ref{eq:decay-anti}) only proceed in one direction, from left to right. If the rates of these decay reactions are not equal, CP violation occurs.
We can then proceed to introduce a CP violation parameter that is adjusted to the total decay rate, which can be expressed as \cite{Davidson:2008bu}
\begin{align}
    \epsilon_1  \equiv \frac{\Gamma_1 - \overline{\Gamma}_1}{\Gamma_1 + \overline{\Gamma}_1}.
\end{align}
The CP violation parameter is nonzero only if the loop corrections are taken into account,
\begin{align}
    \epsilon_1 = \sum_{k\neq1} \frac{1}{8\pi} \frac{\Im \left(yy^{\dagger}\right)_{1k}^2}{\left(yy^{\dagger}\right)_{11}} \left[ f\left(\frac{M_k^2}{M_1^2}\right) + \frac{M_1 M_k}{M_1^2 - M_k^2}\right],
    \label{eq:CP-parameter}
\end{align}
which in that
\begin{align}
    f(x) = \sqrt{x} \left[1-\left(1+x\right)\ln\left(\frac{1+x}{x}\right)\right].
\end{align}

Now, we derive the evolution equation for an RHN with the distribution function $f_{N_1}=f_{N_1}(x^{\alpha},p^{\alpha})$. The classical form of the evolution equation is known as the Boltzmann equation is given by
\begin{align}
    \boldsymbol{L}[f_{N_1}]=\boldsymbol{C}[f_{N_1}]
    \label{eq:Boltzmann},
\end{align}
where $\boldsymbol{L}$ is the Liouville operator, which describes the variation of particles by dynamical parameters, and $\boldsymbol{C}$ is the collision operator, which describes the source of microscopic process evolution.

By perturbation in the metric, we expected the collision operator to not modify and take the standard form, as in Ref. \cite{Kolb:1990vq} because microscopic processes are independent of how they evolve on a large-scale. Although, the Liouville operator is affected which in the relativistic form is given by
\cite{Kolb:1990vq}
\begin{align}
    \boldsymbol{L}=p^{\alpha}\frac{\partial}{\partial x^{\alpha}}-\Gamma^{\alpha}_{\beta \gamma} p^{\beta} p^{\gamma} \frac{\partial}{\partial p^{\alpha}},
    \label{eq:Liouville}
\end{align}
where $\Gamma^{\alpha}_{\beta \gamma}$ are Christoffel symbols of the related metric. For BI metric (\ref{eq:BI-metric}), nonzero Christoffel symbols are equal to
\begin{align}
    \Gamma^{1}_{0 1}=\Gamma^{1}_{1 0}&=\frac{\dot{a_1}}{a_1}, \quad \Gamma^{2}_{0 2}=\Gamma^{2}_{2 0}=\frac{\dot{a_2}}{a_2}, \quad \Gamma^{3}_{0 3}=\Gamma^{3}_{3 0}=\frac{\dot{a_3}}{a_3}, \notag  \\
    &\Gamma^{0}_{1 1}=a_1\dot{a_1}, \quad \Gamma^{0}_{2 2}=a_2\dot{a_2}, \quad \Gamma^{0}_{3 3}=a_3\dot{a_3}.
    \label{eq:Christoffel}
\end{align}
By substituting obtained Christoffel symbols (\ref{eq:Christoffel}) in Liouville operator (\ref{eq:Liouville}), Boltzmann Eq.\ (\ref{eq:Boltzmann}) reduced to
\begin{align}
    \frac{\partial f_{N_1}}{\partial t} - 2 \frac{\dot{a_1}}{a_1} p^1 \frac{\partial f_{N_1}}{\partial p^1} - 2 \frac{\dot{a_2}}{a_2} p^2 \frac{\partial f_{N_1}}{\partial p^2} - 2 \frac{\dot{a_3}}{a_3} p^3 \frac{\partial f_{N_1}}{\partial p^3} = \frac{1}{p^0} \boldsymbol{C}[f_{N_1}].
    \label{eq:anisotropic-Boltzmann}
\end{align}
According to the definition of Hubble rate (\ref{eq:Hubble}) and the definition of the number density of RHN $n_{N_1}=\frac{g_{N_1}}{(2\pi)^3}\int d^3p f$ where $g_{N_1}=2$ is degree of freedom of RHN, Eq.\ (\ref{eq:anisotropic-Boltzmann}) can be given by
\begin{align}
    \frac{dn_{N_1}}{dt}+3Hn_{N_1} = \frac{g_{N_1}}{(2\pi)^3} \int \boldsymbol{C}[f_{N_1}] \frac{d^3p}{p^0}.
\end{align}
Now, according to $sa^3=\rm const.$ where $s$ is entropy, and the definition of $Y_{N_1} \equiv n_{N_1}/s$, we can simplify the last Eq. as
\begin{align}
    \frac{dY_{N_1}}{dt} = \frac{g_{N_1}}{s (2\pi)^3} \int \boldsymbol{C}[f] \frac{d^3p}{p^0}.
\end{align}
Finally, the r.h.s. of Eq. above is substituted from Ref. \cite{Kolb:1990vq} to achieving
\begin{align}
    \frac{dY_{N_1}}{dt} = - 2 \langle\Gamma_1\rangle \left( Y_{N_1} - Y_{N_1}^{\rm eq} \right),
\end{align}
where $Y_{N_1}^{\rm eq}$ is the equilibrium value of the number density of the RHN which is expressed in Eq.\ (\ref{eq:YEq}). By introducing a dimensionless parameter $z \equiv M_1/T$, we can change the variable of the obtained equation as
\begin{align}
    \frac{dY_{N_1}}{dz} = \frac{dt}{dT} \frac{dT}{dz} \left[- 2 \langle \Gamma_1 \rangle \left( Y_{N_1} - Y_{N_1}^{\rm eq} \right)\right].
    \label{eq:anisotropic-Boltzmann-l}
\end{align}

Similarly, Boltzmann equation for $Y_{B-L}\equiv (\overline{n}_l - n_l)/s$ according to the collision term as mentioned in Ref. \cite{Kolb:1990vq} is given by
\begin{align}
    \frac{dY_{B-L}}{dz} = \frac{dt}{dT} \frac{dT}{dz} \left[- \epsilon_1 2 \langle \Gamma_1 \rangle \left( Y_{N_1} - Y_{N_1}^{\rm eq} \right) - \frac{Y_{N_1}^{\rm eq}}{Y_{l}^{\rm eq}} \langle \Gamma_{1} \rangle Y_{B-L}\right],
    \label{eq:anisotropic-Boltzmann-r}
\end{align}
where $Y_{l}^{\rm eq}$ is the equilibrium value of the number density of lepton, which is expressed in Eq.\ (\ref{eq:YEq}).

To obtain $\frac{dt}{dT} \frac{dT}{dz}$, we need a relation between $z$ and $T$ that is well known as $z=M_1/T$ and a relation between $t$ and $T$. 
According to $H = 1/pt$ where $p=2$ for the radiation-dominated universe or $p=3$ for the square of the shear scalar-dominated universe, as mentioned in Sect. \ref{sec:an-isotropic-cosmology}, the relation between $t$ and $T$ could be found as $tT^p=\rm constant$. Then, one can obtain
\begin{align}
    \frac{dt}{dT} \frac{dT}{dz} = \frac{1}{Hz}.
\end{align}

We can simplify Boltzmann equations (\ref{eq:anisotropic-Boltzmann-l}) and (\ref{eq:anisotropic-Boltzmann-r}) as
\begin{align}
    \frac{dY_{N_1}}{dz} &= -D_1 \left( Y_{N_1} - Y_{N_1}^{\rm eq} \right),\\
    \frac{dY_{B-L}}{dz} &= - \epsilon_1 D_1 \left( Y_{N_1} - Y_{N_1}^{\rm eq} \right) - W_1 Y_{B-L},
\end{align}
where the decay parameter $D_1$ and the washout parameter $W_1$ are defined as
\begin{align}
    D_1 \equiv  \frac{2 \langle \Gamma_{1} \rangle}{H z}, \quad
    W_1 \equiv \frac{1}{2} \frac{Y_{N_1}^{\rm eq}}{Y_{l}^{\rm eq}} D_1,
    \label{eq:D1W1}
\end{align}
which in these $Y_{\chi}^{\rm eq}$ denotes the equilibrium value of the number density of $\chi$ is expressed as
\cite{Kolb:1990vq}
\begin{align}
    Y_{N_1}^{\rm eq} = \frac{45}{4 \pi^4} \frac{g_{N_1}}{g_{\star}} z^2 K_2(z), \quad 
    Y_{l}^{\rm eq} \simeq \frac{45}{4 \pi^4} \frac{g_{l}}{g_{\star}} \frac{3}{2} \zeta(3),
    \label{eq:YEq}
\end{align}
where $g_{N_1}=g_{l}=2$ are the degrees of freedom and $\zeta(s)$ is the Riemann zeta function.

\begin{figure}[h]
    \includegraphics[width=140mm]{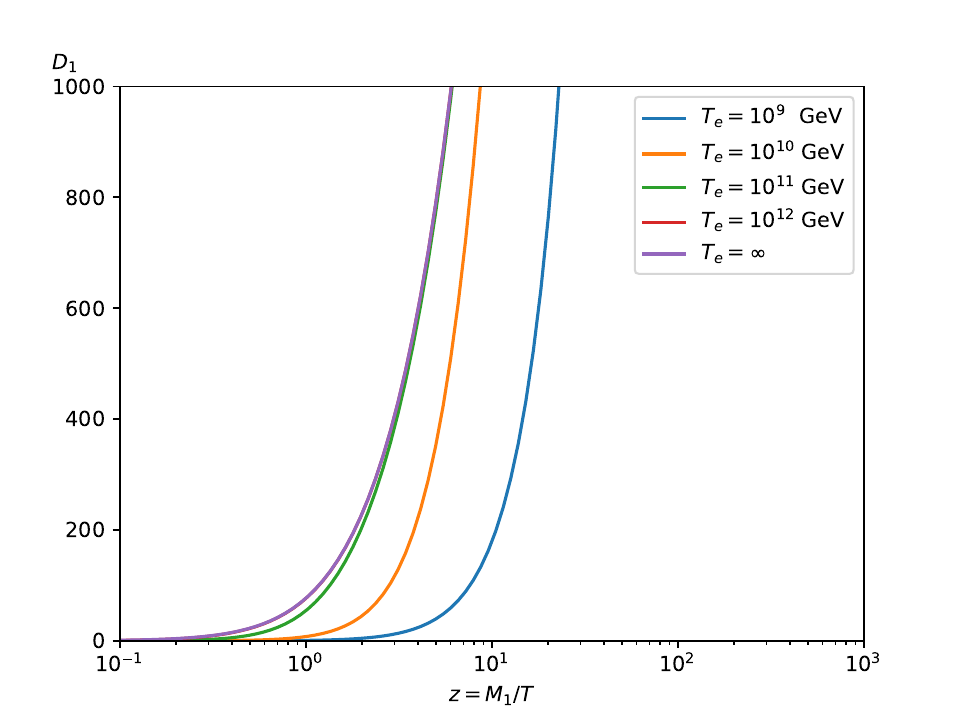}
    \caption{The decay parameters for some $T_e$ values with $M_1 = 10^{11}\ \rm GeV$ \label{fig:D1}}
\end{figure}

According to Eq. (\ref{eq:D1W1}), we plot the evolution of the decay parameter for some values of $T_e$ in Fig.\ \ref{fig:D1}. As one can see, for $T_e<M_1$ cases, the rising decay parameter occurs later than $T_e>M_1$ cases, which did not accept any effect. This implies that generating asymmetry for $T_e<M_1$ cases occurs later in $z$ than in the standard case, but as we work at very high temperatures, this does not affect the asymmetry near the electroweak phase transition.

\begin{figure}[h]
    \includegraphics[width=140mm]{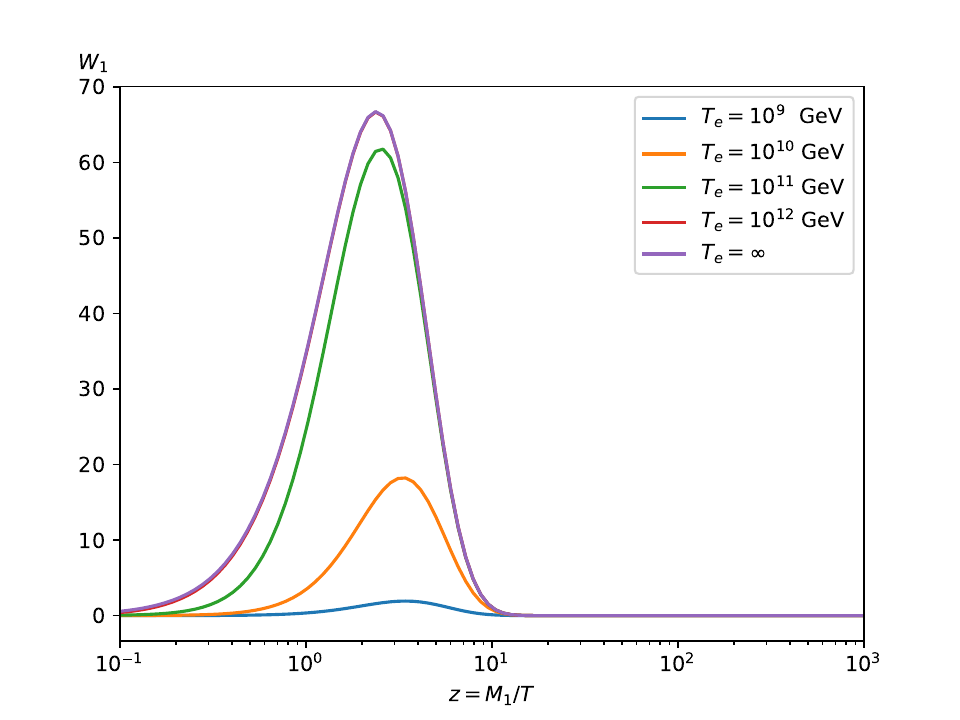}
    \caption{The washout parameters for some $T_e$ values with $M_1 = 10^{11}\ \rm GeV$ \label{fig:W1}}
\end{figure}

Similar to the decay parameter, according to Eq.\ (\ref{eq:D1W1}) we plot the evolution of washout parameter for some $T_e$ values in Fig.\ \ref{fig:W1}. For $T_e<M_1$ cases, in addition to reducing the washout strength, its maximum point moves to lower temperatures. However, by reducing the amount of washout, the movement of the maximum point does not have a significant effect.

The generated $Y_{B-L}$ can be converted to the baryon asymmetry $Y_B$ through the weak sphaleron process. Upon taking into account the hypercharge neutrality condition, the weak and strong sphaleron processes, and all chirality flip processes, the baryon asymmetry is obtained as \cite{Chen:2007fv} 
\begin{align}
    Y_{B} = \frac{28}{79} Y_{B-L}.
    \label{eq:YBLtoB}
\end{align}

\section{Numerical results}
\label{sec:results}
In this section, before we numerically solve the obtained evolution equations, we would parameterize the Yukawa matrix in the Casas-Ibarra way for three RHNs as \cite{Casas:2001sr}
\begin{align}
    y=-iU\sqrt{m} R^T(\omega_1,\omega_2,\omega_3) \sqrt{M} \frac{\sqrt{2}}{v},
    \label{eq:Casas-Ibarra}
\end{align}
where $v=246\ \rm GeV$ denotes the Higgs expectation value.
$m$ is the diagonal mass matrix of light neutrinos and $U$ is the unitary neutrino mixing matrix, known as the PMNS matrix (Pontecorvo-Maki-Nakagawa-Sakata matrix). The PMNS parameters and mass splits of active neutrinos taken from the normal hierarchy for masses are taken from NuFIT 5.2 \cite{Esteban:2020cvm}.
Note that there are also two phases of Majorana: $\alpha_{21}$ and $\alpha_{31}$, which can take values between $0$ and $4\pi$. Here, we neglect the Majorana phases.
Moreover, $M$ is the diagonal matrix of the RHN masses.
Finally, $R(\omega_1,\omega_2,\omega_3)$ is a generic three-dimensional orthogonal complex matrix generated via three complex angles $\omega_i \equiv x_i+iy_i$.
In summary, there are ten free parameters in the theory. They are listed in Tab. \ref{tab:parameters} along with their values considered in this work.
\begin{table}[h]
    \caption{Free parameters of the theory: $m$ is the lightest active neutrino mass, $M_i$ are RHNs masses, $x_i$ and $y_i$ are parameters of the $R$ orthogonal complex matrix \label{tab:parameters}}
    \begin{ruledtabular}
        \begin{tabular}{c c c c c c c c c c} 
            $m/{\rm GeV}$ & $M_1/{\rm GeV}$ & $M_3/{\rm GeV}$	& $M_3/{\rm GeV}$ & $x_1/\degree$ & $y_1/\degree$ & $x_2/\degree$ & $y_2/\degree$ & $x_3/\degree$ &   $y_3/\degree$ \\
            \colrule
            $10^{-11}$	& $10^{11}$ & $10^{11.6}$& $10^{12}$ & $12$ & $51.4$ & $33$ & $11.4$ & $180$ & $11$\\
        \end{tabular}
    \end{ruledtabular}
\end{table}

\begin{figure}[h]
    \includegraphics[width=140mm]{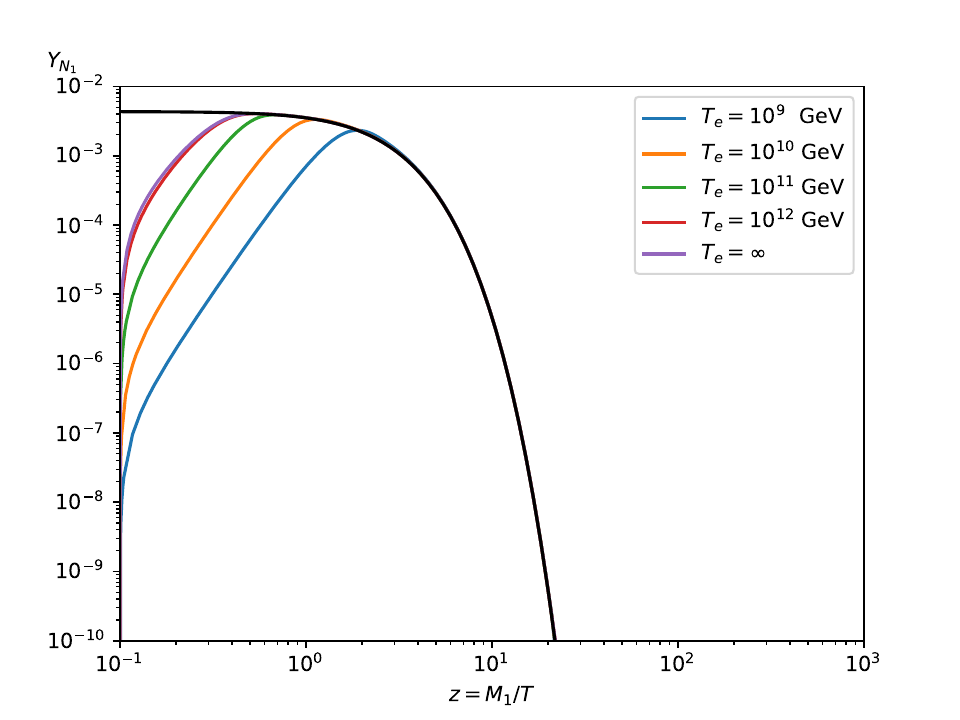}
    \caption{Evolution of $Y_{N_1}$ for some $T_e$ values where the black line is $Y_{N_1}^{\rm eq}$ \label{fig:YN1}}
\end{figure}

\begin{figure}[h]
    \includegraphics[width=140mm]{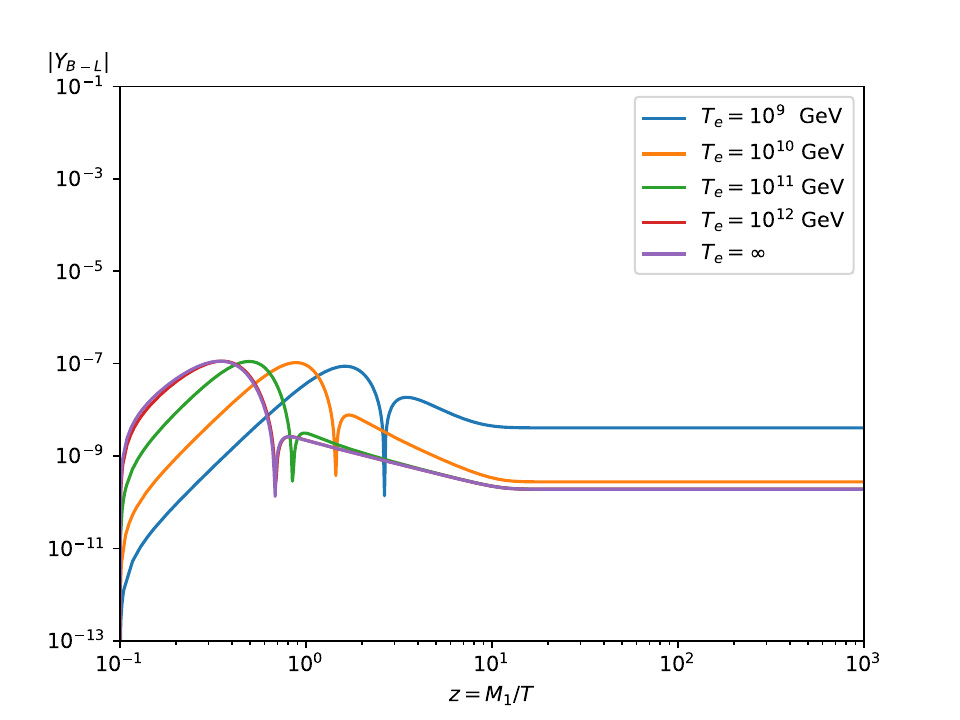}
    \caption{Evolution of $|Y_{B-L}|$ for some $T_e$ values \label{fig:YBL}}
\end{figure}

Now, we want to numerically solve the obtained evolution equations simultaneously from the starting point $z_0=10^{-1}$ to the electroweak phase transition by considering zero initial asymmetries. By solving the evolution equations for certain values of $T_e$, we present the numerical solutions of $Y_{N_1}$ and $Y_{B-L}^q$ in Figs.\ \ref{fig:YN1} and \ref{fig:YBL}, respectively.
By examining the parameter space, it was established that, when $T_e>10^9\ {\rm GeV}$, a strong washout regime prevails, distinguished by $\Gamma_1 > H(T = M_1)$. In this regime, the final result is independent of $Y^q_{N_1}(z_0)$ \cite{Buchmuller:2004nz}. Consequently, we initially assume that $Y_{N_1}^q(z_0)=0$.

As a result, it can be shown in Fig. \ref{fig:YN1} the producing of $Y_{N_1}$ commences at a later time for cases with $T_e<M_1$ compared to those with $T_e>M_1$ that did not experience any effect. Consequently, for $T_e<M_1$ the maximum amount of the the RHN produced is reduced. This leads to a reduction in CP violation through RHN decay, which in turn affects baryon asymmetry. This behavior reduces baryon asymmetry rather than increasing it by decreasing the washout parameter.
It is important to note that for $T_e>10^9\ {\rm GeV}$, this effect is negligible because of the strong washout regime. However, for $T_e<10^9\ {\rm GeV}$, one can this effect can be counteracted by assuming a nonzero initial RHN abundance, for example, that produced by an alternative nonthermal mechanism \cite{Giudice:2003jh}.

In Fig.\ \ref{fig:YBL}, we show obtained $|Y_{B-L}|$ for some $T_e$ values. As we can see, the washout strength for the $T_e<M_1$ cases is weak. The maximum amount of $|Y_{B-L}|$ was reduced, as explained in the previous paragraph. In addition, moving the maximum amount of $Y_{B-L}$ to lower temperatures for $T_e<M_1$ cases is related to the delayed production of RHN, as shown in Fig.\ \ref{fig:D1}.

\begin{figure}[h]
    \includegraphics[width=140mm]{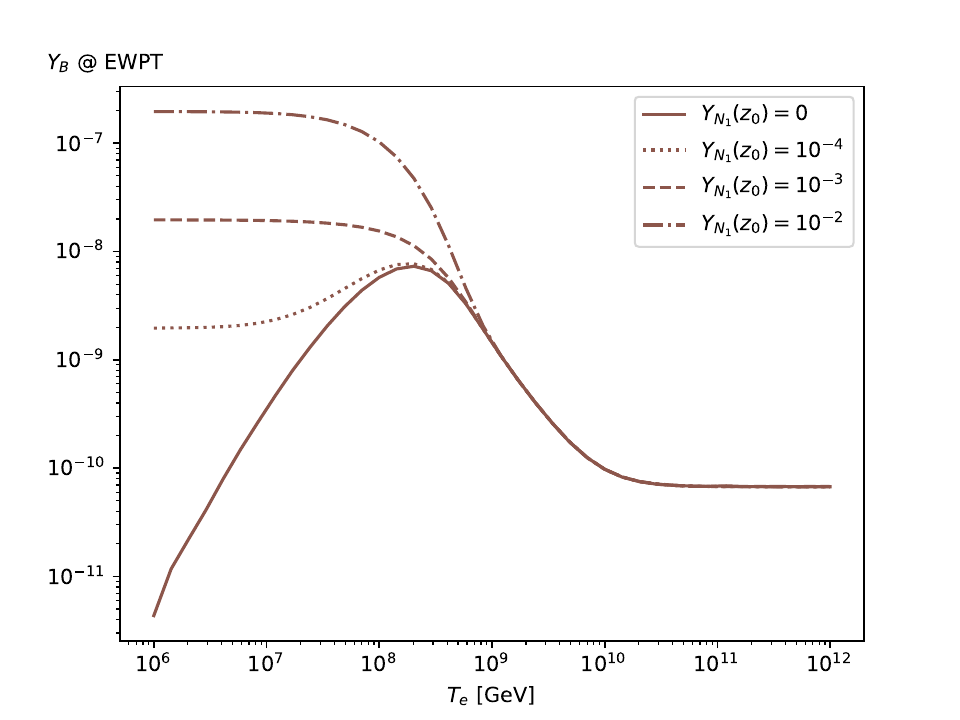}
    \caption{Evolution of $Y_{B}$ at EWPT versus $T_e$ \label{fig:YB}}
\end{figure}

Undertaking an examination of enhancing baryon asymmetry through reducing the washout parameter and diminishing baryon asymmetry via the behavior of $Y_{N_1}$, we have plotted $Y_B$, derivable through the conversion of $Y_{B-L}$ employing Eq.\ (\ref{eq:YBLtoB}), at the electroweak phase transition, versus to the $T_e$ for specific initial abundances of RHN.
As shown in Fig.\ \ref{fig:YB}, there is no impact for $T_e>M_1$ cases. Second, as anticipated, for $T>10^9\ {\rm GeV}$, the final outcome was insensitive to the initial abundance of RHN. Furthermore, it is evident that baryon asymmetry in this anisotropy range can be enhanced. Third, for $T_e<10^9\ {\rm GeV}$, baryon asymmetry is contingent on the initial abundance of RHN and can increase or decrease, as previously discussed.
Consequently, leptogenesis can be probed by reaching a low-scale RHN mass \cite{Chun:2017spz}.

\section{Conclusion}
\label{sec:conclu}
In this study, we examined the effect of the universe anisotropy on thermal leptogenesis.
The study found that anisotropy can affect the production of baryon asymmetry in thermal leptogenesis by modifying the decay and washout parameters.
Indeed, after numerically solving the evolution equations, we found that, for anisotropic cases, the generated baryon asymmetry can be greater than the standard for the specific strength of anisotropy.

Future studies will reveal the validity of isotropic assumptions for the universe. Until then, we could study the effects on any area of the early universe, especially baryogenesis and other leptogenesis scenarios.

\section*{Acknowledgments}
The author is very grateful to S. Safari, and S. S. Gousheh for their helpful comments.

\bibliography{biblio}

\end{document}